\documentclass[sigconf]{acmart}
\AtBeginDocument{%
  }

\copyrightyear{2025}
\acmYear{2025}
\setcopyright{cc}
\setcctype{by-nc-sa}

\acmConference[RecSys '25]{Proceedings of the Nineteenth ACM Conference on Recommender Systems}{September 22--26, 2025}{Prague, Czech
Republic}
\acmBooktitle{Proceedings of the Nineteenth ACM Conference on Recommender Systems (RecSys '25), September 22--26, 2025, Prague, Czech
Republic}\acmDOI{10.1145/3705328.3759302}
\acmISBN{979-8-4007-1364-4/2025/09}

\usepackage{array}
\usepackage{amsmath}
\usepackage{xspace}
\usepackage[ruled,vlined]{algorithm2e}
\newcolumntype{H}{>{\setbox0=\hbox\bgroup}c<{\egroup}@{}}

\newcommand{\itemletter}{t}
\newcommand{\Itemletter}{T}

\newcommand{\eg}{e.g., }
\newcommand{\ie}{i.e., }
\newcommand{\wrt}{w.r.t. }

\newcommand{\ndcg}{\text{NDCG}\xspace}

\newcommand{\coverage}{\text{Cov}\xspace}

\newcommand{\efd}{\text{EFD}\xspace}

\newcommand{\epc}{\text{EPC}\xspace}

\newcommand{\arp}{\text{ARP}\xspace}

\newcommand{\popreo}{\text{PopREO}\xspace}

\newcommand{\poprsp}{\text{PopRSP}\xspace}

\newcommand{\ndcgName}{NDCG}

\newcommand{\covName}{Coverage}
\newcommand{\coverageName}{\covName} 

\newcommand{\metricAt}[2]{\text{#1}@\ensuremath{#2}\xspace}

\newcommand{\ndcgfive}{\metricAt{\ndcgName}{5}}

\newcommand{\deepmf}{\text{DeepMF}\xspace}
\newcommand{\dmf}{\text{DeepMF}\xspace}

\newcommand{\bpr}{\text{BPR}\xspace}

\newcommand{\mlonem}{\text{ML-1M}\xspace}

\newcommand{\amazonbaby}{\text{Baby}\xspace}
\newcommand{\amazonmusic}{\text{Music}\xspace}

\newcommand{\multvae}{\text{MultVAE}\xspace}
\newcommand{\lightgcn}{\text{LightGCN}\xspace}

\newcommand{\cobrarVar}{\texttt{CoBraR}}  
\newcommand{\cobrar}{\cobrarVar\xspace}

\begin{document}

\title[Collaborative Branch for Recommendation]{Parameter-Efficient Single Collaborative Branch for Recommendation}


\author{Marta Moscati}
\email{marta.moscati@jku.at}
\orcid{0000-0002-5541-4919}
\authornote{Contact Author.}
\affiliation{%
  \institution{Institute of Computational Perception, Johannes Kepler University Linz}
  \streetaddress{Altenberger Straße 69}
  \city{Linz}
  \country{Austria}
}

\author{Shah Nawaz}
\email{shah.nawaz@jku.at}
\orcid{0000-0002-7715-4409}
\affiliation{%
  \institution{Institute of Computational Perception, Johannes Kepler University Linz}
  \streetaddress{Altenberger Straße 69}
  \city{Linz}
  \country{Austria}
}

\author{Markus Schedl}
\email{markus.schedl@jku.at}
\orcid{0000-0003-1706-3406}
\affiliation{
  \institution{Institute of Computational Perception, Johannes Kepler University Linz and AI Lab, Linz Institute of Technology}
  \city{Linz}
  \country{Austria}
}

\newcommand{\embdim}{N}
\newcommand{\cluster}{cluster}
\newcommand{\clusters}{clusters}
\newcommand{\Cluster}{Cluster}
\newcommand{\Clusters}{Clusters}

\newif\ifworkinprogress
\workinprogressfalse

\ifworkinprogress
	\newcommand{\ms}[1]{\textcolor{blue}{{[Markus] #1}}}
	\newcommand{\mm}[1]{\textcolor{olive}{{[Marta] #1}}}
	\newcommand{\sn}[1]{\textcolor{green}{{[Shah] #1}}}
  \newcommand{\newtext}[1]{\textcolor{purple}{#1}}    
\else
    \newcommand{\ms}[1]{}
    \newcommand{\mm}[1]{}
    \newcommand{\sn}[1]{}
    \newcommand{\newtext}[1]{#1}
\fi

\renewcommand{\shortauthors}{Moscati et al.}

\begin{abstract}
  Recommender Systems (RS) often rely on representations of users and items in a joint embedding space and on a similarity metric to compute relevance scores. In modern RS, the modules to obtain user and item representations consist of two distinct and separate neural networks (NN). In multimodal representation learning, weight sharing has been proven effective in reducing the distance between multiple modalities of a same item. Inspired by these approaches, we propose a novel RS that leverages weight sharing between the user \textit{and} item NN modules used to obtain the latent representations in the shared embedding space.
  The proposed framework consists of a single \textbf{Co}llaborative \textbf{Bra}nch for \textbf{R}ecommendation (\cobrar{}). We evaluate \cobrar{} by means of quantitative experiments on e-commerce and movie recommendation. 
  Our experiments show that 
  by reducing the number of parameters and improving beyond-accuracy aspects without compromising accuracy, \cobrar{} has the potential to be applied and extended for real-world scenarios.
\end{abstract}

\ccsdesc[500]{Information systems~Recommender systems}
\ccsdesc[300]{Computing methodologies~Neural networks}

\keywords{Recommender Systems, Single-Branch Network, Weight Sharing, Beyond-Accuracy Evaluation, Collaborative Filtering, Multimedia Recommendation}

\maketitle

\section{Introduction}
    Most Recommender Systems (RS) 
    assign user--item pairs recommendation scores 
    by representing users and items as vector embeddings in a latent factor space. 
    Modern approaches rely on neural networks (NN) to obtain the vector embeddings~\cite{zhang2022dlrs_rshb}. The architecture of these models typically consists of two modules or branches~\cite{xue2017deepmf,he2017neumf,he2018neumf_outer,melchiorre2022protomf}: one branch encodes the user and one branch encodes the item. The paradigm underlying these models is 
    that for positive user--item interactions, the user and item embeddings are close 
    in the latent factor space 
    and, therefore, result in a high recommendation score. This paradigm resembles approaches in multimodal representation learning~\cite{baltruvsaitis2018multimodal,multimodal_survey,xu2023multimodal,feng2020deep}, where embeddings of multiple \newtext{heterogeneous} modalities 
    can be used simultaneously or interchangeably to perform a task (\eg identify a person from their face and/or voice)
    . Although multi-branch architectures~\cite{wang2016learning,nagrani2018learnable,lu2019vilbert,radford2021clip,saeed2022fusion,hannan2025paeff} are standard in multimodal representation learning, recently, single-branch architectures 
    have been emerging as effective alternative \newtext{to learn joint representations from heterogeneous modalities}~\cite{tschannen2023clippo,saeed2023single,liaqat2025chameleon}.
    These architectures share the same NN for more than one modality and offer the advantage of a reduced number of model parameters~\cite{tschannen2023clippo}. Single-branch architectures have been recently applied in the domain of multimodal RS to tackle cold-start scenarios~\cite{ganhoer_moscati2024sibrar}. However, \newtext{the architecture of these multimodal RS} still relies on two separate single-branch encoders - one for the user and one for the item. Therefore, the question of whether NN-based RS can effectively leverage single-branch architectures to encode users and items \emph{with a same module} remains unanswered. To fill this gap, in this paper we propose the use of a single \textbf{Co}llaborative \textbf{Bra}nch for \textbf{R}ecommendation (\cobrar{}). To investigate the effectiveness of single-branch architectures in recommendation we ask the following research questions (RQ): \begin{description}
        \item[RQ1: ]What is the performance of single-branch RS compared to two-branch RS in terms of recommendation quality? 
        \item[RQ2: ]How does the reduction in number of parameters impact the recommendation accuracy of single-branch architectures? 
    \end{description}
    We answer these questions by means of quantitative experiments on datasets from the domains of e-commenrce and movie recommendation
    . Our analysis shows that \cobrar{} reaches comparable or superior recommendation accuracy than two-branch architectures, with substantial improvement on beyond-accuracy aspects and a reduced number of model parameters. Our work is the first to leverage single-branch networks only on collaborative data, therefore establishing a new paradigm for collaborative filtering RS. 
\section{Background and Related Work}
Latent factor models are very common for recommendation~\cite{koren2009matrix,koren2022collaborative_filtering_handbook}. These approaches aim at obtaining latent user and item representations that are reflective of the interaction data. Based on these representations, for each user--item pair the model predicts a recommendation score quantifying the model's confidence that the user will interact with the item. Since the score is typically obtained via inner product or cosine similarity, these approaches aim at obtaining 
representations that are close to each other in the shared embedding space for  positive user--item pairs, and further apart for negative ones.  
Similarly, multimodal learning~\cite{baltruvsaitis2018multimodal,xu2023multimodal,feng2020deep} aims to obtain latent representations of multiple data sources (\eg text, images, audio) that are close to each other in the embedding space for semantically related samples (\eg lyrics, album cover, and audio signals of music tracks of a same artist), and further apart for unrelated ones. Multi-branch architectures, in which each modality is processed by a separate branch, are standard in multimodal learning~\cite{manzoor2023multimodality}. 
Only recently single-branch architectures, in which more than one modality is processed by the same branch, have been emerging as effective alternatives to multi-branch ones. These approaches effectively combine multimodal data in a shared latent embedding space~\cite{nawaz2019deep,nawaz2019cross,saeed2022fusion}, with the additional advantage of a reduction in number of model parameters~\cite{tschannen2023clippo}. 
The only application of the single-branch paradigm to recommendation is  SiBraR~\cite{ganhoer_moscati2024sibrar}, in which the user and item multimodal data are mapped to the shared embedding space by \textit{two} distinct single-branches
. 
Therefore, to our knowledge we are the first to use only \textit{one} single-branch to obtain both user and item latent representations.

\section{Methodology}
    \label{sec:methodology}
    \textit{Preliminaries. } We denote the set of $N$ users with $\mathcal{U} = \{u_i\}_{i=1}^{N}$ and the set of $M$ items with $\mathcal{T} = \{\itemletter_j\}_{j=1}^{M}$. We consider a scenario of implicit feedback and denote with $\mathcal{I} = \{(u_i, \itemletter_j)\}$ the user--item pairs corresponding to a positive feedback; those are collected as non-zero entries in the binary interaction matrix $R \in \mathbb{R}^{N\times M}$.
    
\begin{figure}[t]
        \centering
            \includegraphics[width=0.48\textwidth]{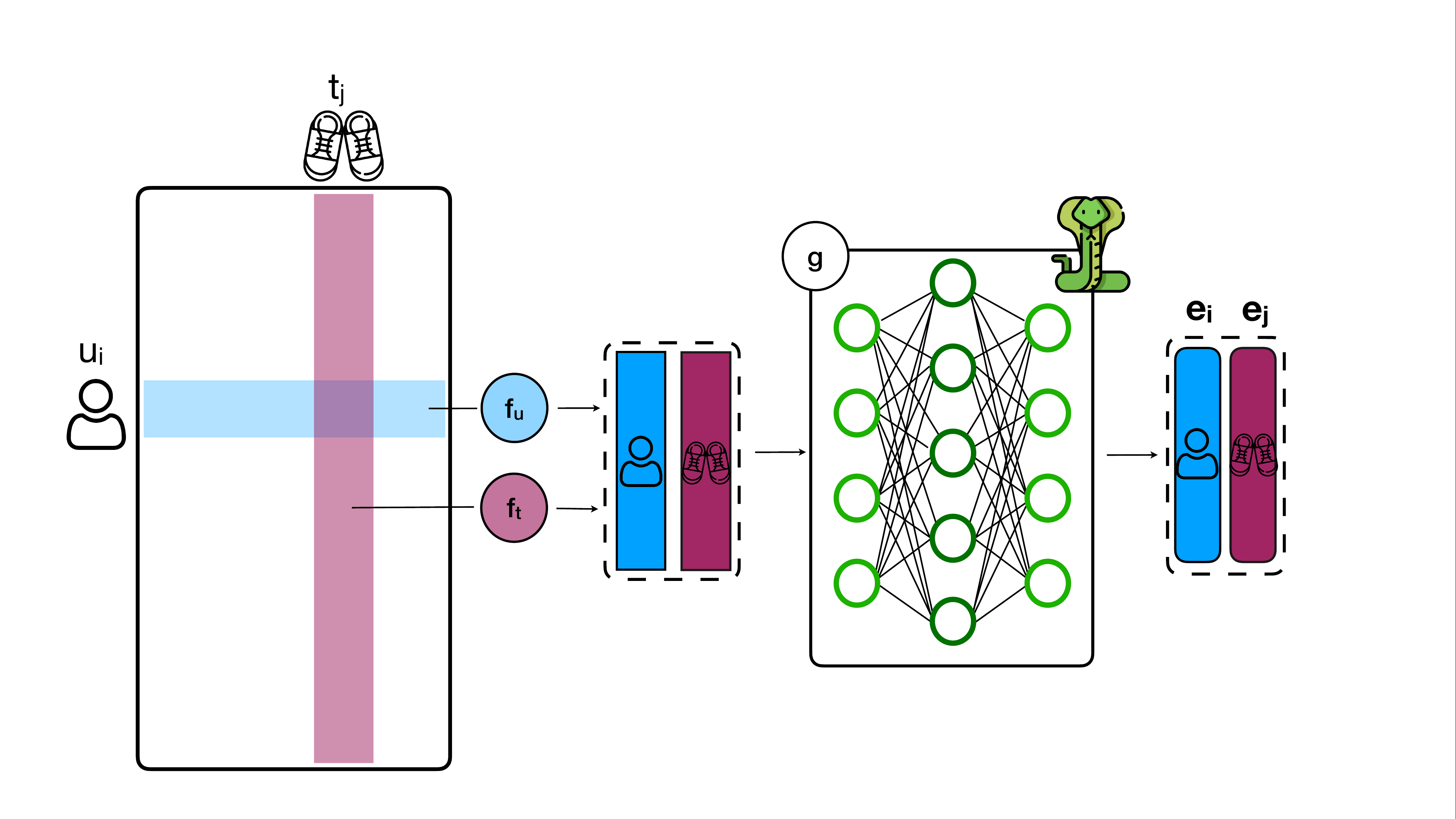}
            \caption{Architecture of \cobrar{}.  The left block represents the user--item interaction matrix $R\in \mathbf{R}^{N\times M}$. $f_{u}$ and $f_{\itemletter}$ represent the single fully-connected linear layers downprojecting the rows and columns of $R$ to dense vectors
            , while $g$ represents the multi-layer  collaborative branch shared between users and items. For each user--item interaction pair $(u_i, \itemletter_j)$, \cobrar{} computes the recommendation score as cosine similarity between the user and item embeddings, $\hat{y}_{ij} = \cos{(\mathbf{e_{i}}, \mathbf{e_{j}})}$.}
            \label{fig:cobrar}
    \end{figure}

    \textit{Recommendation Score. }  Figure~\ref{fig:cobrar} gives an overview of \cobrar{}'s architecture. Given a user--item pair $(u_i, \itemletter_j)$, \cobrar{} outputs a recommendation score as described in the following. Analogously to DeepMF~\cite{xue2017deepmf}, user $u_i$ is initially represented as $R_{i*} \in \mathbb{R}^{M}$, \ie the $i$-th row of the interaction matrix. This initial sparse vector is down-projected to a dense $p$-dimensional\footnote{The parameter $p$ is one of the hyperparameters defining the architecture of \cobrar{}. The optimization of all hyperparameters is described in Section~\ref{sec:experiments}.} vector $f_u(R_{i*})\in \mathbb{R}^{p}$ with a fully connected layer. This vector is the user input to the collaborative branch $g$, consisting of several fully connected layers with ReLU as non-linear activation function. The final user embedding is hence given by $\mathbf{e_{i}}=g(f_u(R_{i*}))$. Similarly, the initial representation $R_{*j} \in \mathbb{R}^{N}$ of item $\itemletter_j$ is first down-projected to a dense vector $f_\itemletter(R_{*j})\in\mathbb{R}^{p}$, and then passed to the \textit{same} collaborative branch $g$. The final embedding of item $\itemletter_j$ is hence given by $\mathbf{e_{j}}=g(f_\itemletter(R_{*j}))$. As in \deepmf{}~\cite{xue2017deepmf}, the cosine similarity between the user and item embeddings $\hat{y}_{ij} = \cos{(\mathbf{e_{i}}, \mathbf{e_{j}})}$ is the final recommendation score.
    
    \textit{Training. }For each positive user--item pair $ \{(u_i, \itemletter_j) | R_{ij}=1\}$ we sample $n_\text{neg}$ negative items for user $u_i$ and compute the recommendation scores for the positive and negative items. Analogously to DeepMF~\cite{xue2017deepmf}, we apply a lower threshold $\mu$ to all the scores during training. Then, for the positive pair and all negative pairs, we compute the cross-entropy loss $\mathcal{L} = - \ln{\hat{y}_{ij}} - \Sigma_{k=1}^{n_\text{neg}}\ln(1 - {\hat{y}_{ik}})$. The pseudocode describing the computation of the batch loss in \cobrar{} is summarized in Algorithm~\ref{alg:batchloss}. 
        \input{algorithms/cobrar}
    \begin{algorithm}
        \DontPrintSemicolon
        \caption{\cobrar{}'s pseudocode for batch loss.}\label{alg:batchloss}
            \SetKwInOut{Input}{Input}            \SetKwInOut{Output}{Output}
            \SetKwFunction{negSamples}{negSamp}
            \SetKwFunction{randomSampling}{randomSampling}
            \Input{
                $b \subset \{(u_i, \itemletter_j) | R_{ij}=1\}$: Batch of user--item training interactions\\
                $n_\text{neg}$: Number of negatives per interaction\\
            }
            \Output{$\mathcal{L}$: Loss for the batch
            }
                $\mathcal{L} \leftarrow 0$ \tcp*[l]{Reset loss}
                \For{$(u_i, \itemletter_j)\in b$}{
                    $\mathbf{e_{i}} \leftarrow  g(f_u(R_{i*}))$\tcp*[l]{Embed $u_i$ with \cobrar{}}
                    $\mathbf{e_{j}} \leftarrow  g(f_\itemletter(R_{*j}))$\tcp*[l]{Embed $\itemletter_j$ with \cobrar{}}
                    $\hat{y}_{ij} \leftarrow \cos{(\mathbf{e_{i}}, \mathbf{e_{j}})}$\tcp*[l]{Compute positive logit}             
                    $\mathcal{L} \leftarrow \mathcal{L} - \ln{\hat{y}_{ij}}$ \tcp*[l]{Accumulate loss}

                    $\mathcal{N}_{ij}\leftarrow$ \negSamples{$u_i, \itemletter_j, R$}\tcp*[l]{Sample negatives}
                    \For{$\itemletter_k \in \mathcal{N}_{ij}$}{
                        $\mathbf{e_{k}} \leftarrow  g(f_\itemletter(R_{*k}))$\tcp*[l]{Embed $\itemletter_k$ with \cobrar{}}
                        $\hat{y}_{ik} \leftarrow \cos{(\mathbf{e_{i}}, \mathbf{e_{k}})}$\tcp*[l]{Compute positive logit}             
                        $\mathcal{L} \leftarrow \mathcal{L}- \ln{(1 - \hat{y}_{ij})}$ \tcp*[l]{Accumulate loss}
                    }
                }
    \end{algorithm}
    
    \textit{Inference time. }For each user $u_i$, we rank items $\itemletter_j\in \mathcal{\Itemletter}$ in descending order of $\hat{y}_{ij}$, excluding only those items such that  $R_{ij}=1$, \ie excluding items already interacted with. The ordered list of top-$k$ items constitutes the recommendation list for user $u_i$. 
    
\section{Experimental Setup}
    \label{sec:experiments}
    \textit{Datasets. } We conduct our experiments on \newtext{three datasets}. For movie recommendation we use MovieLens 1M (\mlonem)\footnote{\url{https://grouplens.org/datasets/movielens/1m/}}~\cite{movielens}. This dataset provides user ratings of movies gathered from the MovieLens website.\footnote{\url{https://movielens.org/}} For online shopping we use Amazon Baby products (\amazonbaby) and Amazon Music (\amazonmusic),\footnote{\url{https://cseweb.ucsd.edu/~jmcauley/datasets/amazon_v2/}} \newtext{two subsets} of the larger Amazon Reviews'18 dataset~\cite{ni2019amazonreviews}. We consider an implicit feedback setting where user--item interactions are $1$ if the user interacted with the item and $0$ otherwise. We apply 5-core filtering and a user-based random splitting into $70\%, 10\%, 20\%$ interactions, respectively, for train, validation, and testing. The characteristics of the datasets after pre-processing are reported in Table~\ref{tab:interaction_data}.
    \begin{table}[]
        \begin{tabular}{r|r|r| r}
        Dataset Name &\# Users $u$ & \# Items $m$ & \# Interactions \\ \hline
        \mlonem & 6{,}040 & 3{,}416 & 999{,}611 \\
        \amazonbaby & 19{,}181 &  6{,}366  & 138{,}821 \\
        \amazonmusic & 5{,}082 & 2{,}338 & 30{,}623\\ 
        \end{tabular}
        \caption{Characteristics of the datasets after pre-processing.}
        \label{tab:interaction_data}
    \end{table}

    \textit{Model evaluation. } We measure the quality of recommendation lists (RQ1) in terms of both accuracy and beyond-accuracy aspects. Accuracy is measured in terms of Normalized Discounted Cumulative Gain (\ndcgName). As beyond-accuracy metrics, we measure the Average Recommendation Popularity (\arp)~\cite{yin2012arp} of recommended items, defining popularity in terms of the number of interactions in the training set. We measure catalog coverage (\coverageName) as the percentage  
    of catalog items that appear at least once in the top-$k$ recommendation lists. Finally, we measure the popularity bias of recommendation lists by first grouping items into short-head and long-tail, accounting for a cumulative $80\%$ and $20\%$ of the training interactions, respectively. 
    We compute popularity bias as Popularity-based Ranking-based Statistical Parity (\poprsp)~\cite{zhu2020poprsppopreo}, for which lower values indicate recommendations that are less biased towards short-head, popular items. 
    All metrics are computed over lists of $k$ recommendations. For the user-level metrics (\ndcgName{} and \arp) we  average the values over all users and compute the statistical significance of the difference between the best performing model and the remaining ones by carrying out paired $t$-tests of the distributions over users. We consider a difference to be statistically significant if $p<0.05$ after Bonferroni correction, accounting for the multiple pairwise comparisons. We do not carry out tests of statistical significance for the metrics computed over all recommendation lists globally (\poprsp, \coverageName) since for these metrics there is no distribution of values over users. To show that \cobrar{} allows for a reduction in number of parameters (RQ2) with respect to its two-branch counterpart \dmf{}, we compare the \ndcgName{} of both models for a fixed architecture (\ie number and structure of hidden layers, and embedding dimension) of the encoding branches\newtext{, carrying out tests of statistical significance similarly to those described above}. Let us consider an embedding branch consisting of $L$ layers $[d_1, \dots, d_L]$, where $d_i$ is the number of nodes of the layer and $d_L$ is the final embedding dimension. Both architectures will require a downprojection layer from the user/item profile to the dimension $d_1$ of the first layer. Then, after the downprojection and neglecting the biases, the parameters of the embedding branches of \dmf{} will be $2\cdot \Sigma_{i=1}^{L-1} d_i \times d_{i+1}$, while for \cobrar{} $\Sigma_{i=1}^{L-1} d_i \times d_{i+1}$.

    \textit{Baselines. } The natural baseline for testing \cobrar{}'s use of a single-branch architecture is its two-branch counterpart \deepmf{}. 
    \newtext{To compare \cobrar{}'s performance with RS often used for benchmarking, we include  matrix factorization optimized with Bayesian Personalized Ranking (\bpr{})~\cite{bpr}, MultVAE~\cite{multvae} and \lightgcn{}~\cite{he2020lightgcn}. These models also serve the purpose of covering model classes beyond two-branch DNNs, i.e., autoencoder-based and graph-based, respectively. Although MultVAE and LightGCN are often considered as stronger models than DeepMF and BPR, developing CoBraR on top of MultVAE or LightGCN is not possible, since these models are not based on two-branch architectures.}
    
    \textit{Hyperparameters. } All hyperparameters shared among models are optimized with a grid search over the same values. In particular, we vary the embedding dimension in $\{64, 128\}$, the learning rate in $\{10^{-6}, 10^{-7}\}$, fix the batch size to $256$, and use ADAM optimizer with $L^2$ regularization weight in $\{10^{-2}, 10^{-3}\}$. For \cobrar{} and \deepmf{} we set the negative sampling ratio to $5$. We consider two configurations: a shallow one, consisting of a single intermediate layer of size in $\{ 2048 , 1024 , 512 , 256 \}$ and a deep one, with four intermediate layers of dimensions $ [512, 512, 256, 256]$. \newtext{The latter is indicated as ``deep'' in Figure~\ref{fig:boxplot}.} We use ReLU as non-linear activation function. For \cobrar{} we apply dropout with a rate in $\{0.1, 0.5, 0.9\}$. We train models for a maximum of $100$ epochs, apply early stopping with a patience of $10$, reporting the test-set performance of the model reaching the highest $\ndcgfive$ on the validation set.

    For reproducibility, we implement \cobrar{} in the RS framework Elliot~\cite{anelli2021elliot}, and provide the code and link to pre-processed datasets at \url{https://github.com/hcai-mms/cobrar}.

\section{Results}
    \label{sec:results}
    \subsection{RQ1 - Recommendation quality}
        Table~\ref{tab:rec_quality} shows the values of the evaluation metrics computed on the test set for lists of $k=5$ recommended items. For each dataset and each model, we report the test-set performance of the hyperparameter configuration achieving the highest \ndcgfive on the validation set. For user-level metrics (\ndcgfive, \arp) the superscript $^*$ indicates a p-value $p<0.05$ after taking into account repeated pairwise comparisons with Bonferroni correction. We do not carry out tests of statistical significance for the metrics computed over all recommendation lists globally (\poprsp, \coverageName) since for these metrics there is no distribution of values over users. We first observe that \newtext{LightGCN outperforms DeepMF and CoBraR on all datasets in terms of \ndcgfive; this is not surprising since LightGCN is often considered among the most accurate RS. } 
        \newtext{Our experiments are aimed at showing that \cobrar{} can halve the number of model parameters with respect to \deepmf{} without compromising recommendation accuracy. With that respect, on two out of three datasets (\mlonem, \amazonmusic), \cobrar{} outperforms its two-branch counterpart \deepmf{}, showing its ability to reach a higher accuracy with a simpler architecture. 
        The performance decreases only on \amazonbaby, and in general \cobrar{} demonstrates a competitive \ndcgfive{} \wrt \dmf{}. 
        Furthermore, on all datasets \cobrar{} provides recommendations that have a better coverage of the item catalog (\coverage) than all other baselines. In terms of popularity bias (\poprsp), \cobrar{} outperforms other RS on \amazonbaby and \amazonmusic, and is outperformed by a small margin by its two-branch counterpart \deepmf{} on \mlonem. Finally, \cobrar{}'s recommendations on \amazonmusic are less prone to popularity bias (\poprsp). \cobrar{}'s popularity bias is outperformed by \bpr{} on \amazonbaby and by \deepmf{} on \mlonem.} 
        Overall, these results indicate that the use of weight-sharing allows \cobrar{} to reach a better balance between accuracy and beyond-accuracy aspects of recommendation quality with a single-branch, simplified architecture.
        \begin{table*}[]
            \begin{tabular}{r|rHHrHrr|rHHrHrr|rHHrHrr}
            Model & \multicolumn{7}{c|}{\mlonem} & \multicolumn{7}{c}{\amazonbaby} & \multicolumn{7}{c}{\amazonmusic} \\ 
            & \ndcg{} $\uparrow$ & \efd $\uparrow$  & \epc $\uparrow$  & \arp $\downarrow$  & \popreo $\downarrow$ & \poprsp $\downarrow$ & \coverage $\uparrow$ & \ndcg{} $\uparrow$ & \efd $\uparrow$  & \epc $\uparrow$  & \arp $\downarrow$  & \popreo $\downarrow$ & \poprsp $\downarrow$ & \coverage $\uparrow$ 
            & \ndcg{} $\uparrow$ & \efd $\uparrow$  & \epc $\uparrow$  & \arp $\downarrow$  & \popreo $\downarrow$ & \poprsp $\downarrow$ & \coverage $\uparrow$\\ \hline
            \bpr{}~\cite{bpr} & 0.2531	 &3.2259 &	0.2710 &	1258.5 &	0.9493 &	0.9882 &	28.9\%  & 0.0113 & 	 & 	 & $^{*}\mathbf{111.0}$	 & 	 	 &0.9886 & $38.6\%$  & $^{*}\mathbf{0.1044}$	& 0.5575		& 0.0587	& 	43.79	& 	0.7981	& 	0.8247	& 	0.8\% \\
            \dmf{}~\cite{xue2017deepmf}  & 0.2301 &	2.8914 &	0.2463 &	$^{*}\mathbf{1038.4}$ &	0.9506 &	$\mathbf{0.9747}$ &	32.0\% & 0.0156 & $\mathbf{0.0623}$ & $^{*}\mathbf{0.00782}$& 323.5 & 1.0000 & 1.0000 & $1.9\%$& 0.0245	 & 0.1269 & 	0.0140 & 	54.86 & 	1.0000 & 	0.9875 & 	34.6\%\\
            \multvae{}\cite{multvae} & 0.2262 &	2.9617 &	0.2507 &	1271.2 &	0.9367 &	0.9889 &	27.0\% & 0.0155 & & & 320.7& & 1.000 & 3.0\% & 0.0973 & 	0.5278 & 	0.0552 & 	44.10 & 	0.8898 & 	0.9371 & 	55.1\%\\
            \lightgcn{}~\cite{he2020lightgcn}& $^{*}\mathbf{0.2726}$	 &3.3380 &	0.2787 &	1362.1 &	0.9582 &	0.9904 &	28.0\%  & $^{*}\mathbf{0.0252}$ & & & 197.4 & & 0.9579 & 51.0\% & 0.0855	 & 0.4352 & 	0.0482 & 	69.89 & 	0.9085 & 	0.9570 & 	38.1\%\\
            \cobrar{}  & 0.2431 &	3.0204 &	0.2579 &	1056.1 &	0.9354 &	0.9750 &	$\mathbf{35.9}$\% & 0.0124 & $\mathbf{0.0623}$ & 0.00662 & 124.7 & $\mathbf{0.8749}$ & $\mathbf{0.8432}$ & $\mathbf{82.8\%}$ & 0.0397 & 	0.2173 & 	0.0234 & 	$^{*}\mathbf{40.72}$ & 	0.9160 & 	$\mathbf{0.7757}$ & 	$\mathbf{82.6\%}$\\
            \end{tabular}
            \caption{Test-set performance of the hyperparameter configuration achieving the highest \ndcgfive on the validation set. Bold indicates the best performing model. For user-level metrics (\ndcgName, \arp) the superscript $^*$ indicates $p<0.05$ after Bonferroni correction. We do not carry out tests of statistical significance for the metrics computed over all recommendation lists globally (\poprsp, \coverageName) since for these metrics there is no distribution of values over users.}
            \label{tab:rec_quality}
        \end{table*}
        
    \subsection{RQ2 - Accuracy vs. number of parameters}
\begin{figure}
            \centering
                \includegraphics[width=0.48\textwidth]{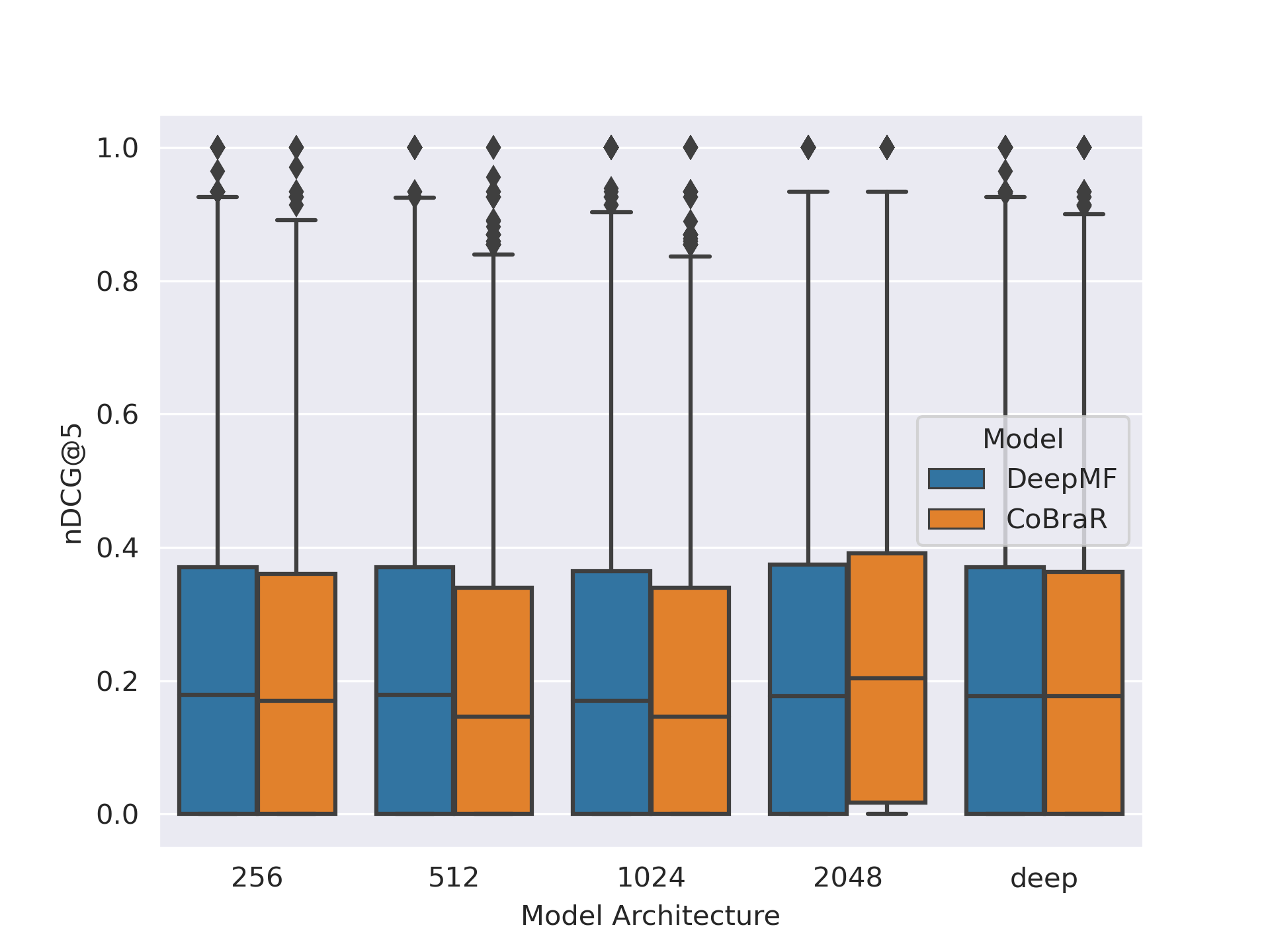}
                \caption{Boxplots summarizing the distributions of test \ndcgfive over users for \mlonem. Left/Blue: \deepmf{}, Right/Yellow: \cobrar{}. }
                \label{fig:boxplot}
            \end{figure}

        To show that \cobrar{} allows for a reduction in the number of parameters without substantially affecting the accuracy of recommendations, we compare the performance of \cobrar{} with its two-branch counterpart \deepmf{} on \mlonem. For \deepmf{} we vary the architecture of the user and item embedding networks, for \cobrar{} we vary the architecture of the collaborative branch. We pair the \deepmf{} and \cobrar{} model instances for which these architectures are equal. We fix the final embedding dimension to $128$, since both models showed a better performance with this value in the overall hyperparameter optimization. All other hyperparameters (learning rate, regularization weight, dropout rate) are fixed separately to the values identified in the hyperparameter optimization. In this way, for each pair of models, the user and item embedding modules of \deepmf{} correspond to twice the number of parameters of the collaborative branch of \cobrar{}.  Figure~\ref{fig:boxplot} shows the boxplots summarizing the distributions of test \ndcgfive over users, for pairs of models of fixed architectures, and with architectures varying over the same values of the hyperparameter optimization. The left, blue boxplots correspond to \deepmf{} and the right, yellow ones to \cobrar{}. 
    The highest accuracy is achieved by \cobrar{} with a shallow architecture with one hidden layer of dimension $2{,}048$, corresponding to the configuration identified in the overall hyperparameter optimization. For smaller architectures (one intermediate layer of dimension $256, 512$, or $1{,}024$) \deepmf{} outperforms \cobrar{}. However overall, and in particular for a deeper, larger architecture consisting of four hidden layers of dimensions $ [512, 512, 256, 256]$, the distributions do not differ substantially. This indicates that, despite halvening the number of parameters for the user and item embedding modules, \cobrar{} accuracy is comparable to the performance of its two-branch counterpart.

    \section{Conclusions, Limitations, and Future Work}
    \label{sec:conclusions}
    In this work we proposed \cobrar{}, a novel collaborative filtering algorithm that, inspired by multimodal representation learning, uses weight sharing to encode both users and items with a same single-branch module. Our experiments showed that \cobrar{} allows to halve the number of parameters and simultaneously reach a better balance between accuracy and beyond-accuracy aspects of recommendation quality, compared to existing two-branch RS. By reducing the number of parameters without compromising recommendation accuracy and simultaneously improving beyond-accuracy qualities, our algorithm has the potential to be applied and extended for real-world, in-production scenarios.

    There are several limitations to our work. First, we limited our experiments to a version of \cobrar{} inspired by \deepmf{}. We therefore used cross-entropy as point-wise loss function during training. We did not investigate the use of other loss functions. In particular, other recommendation losses such as the pairwise BPR loss, more closely resemble the contrastive losses used in multimodal representation learning. Since multimodal representation learning inspired the development of \cobrar{}, these losses have the potential to further improve \cobrar{}'s performance. Analogously, as done in \deepmf{}, we relied on cosine similarity as function for aggregating user and item embeddings to a final recommendation score. 
    We did not investigate the use of more sophisticated aggregation techniques, such as those leveraged in NeuMF~\cite{he2017neumf}. 
    Furthermore, we developed \cobrar{} as a collaborative filtering algorithm, therefore leveraging only user--item interaction data. However, \cobrar{} could be extended to additionally incorporate user and item side-information, \ie combining the two branches of SiBraR~\cite{ganhoer_moscati2024sibrar} into a single one. Finally, we showed how \cobrar{} allows for a reduction in the number of parameters. Since previous works~\cite{desai2022size_tradeoffs} showed that reducing model size comes at the cost at a longer training time, the question of whether \cobrar{} requires a longer training time compared to equivalent two-branch architectures, remains unanswered. We leave these extensions of our work for future research.

\begin{acks}
    This research was funded in whole or in part by the Austrian Science Fund (FWF) \url{https://doi.org/10.55776/P33526}, \url{https://doi.org/10.55776/DFH23}, \url{https://doi.org/10.55776/COE12}, \url{https://doi.org/10.55776/P36413}. 

\end{acks}
\newpage
\balance
\bibliographystyle{ACM-Reference-Format}
\bibliography{main.bib}
\end{document}